\documentclass{emulateapj}
\usepackage{apjfonts}
\usepackage{epsfig}

\journalinfo{The Astrophysical Journal, in press}
\slugcomment{Received 2009 August 4; accepted 2009 August 17}

\shortauthors{CAMILO ET AL.}
\shorttitle{PSR~J1437--5959 in SNR~G315.9--0.0}

\begin{document}

%  Definitions
\def\ergs{erg\,s$^{-1}$}
\def\pccc{pc\,cm$^{-3}$}
\def\kms{km\,s$^{-1}$}
\def\rlum{mJy\,kpc$^2$}
\def\swift{{\em Swift}}

% Objects
\def\psr{PSR~J1437--5959}
\def\snr{G315.9--0.0}

\title{Out of the frying pan: a young pulsar with a long radio trail
emerging from SNR~G315.9--0.0 }

\author{F.~Camilo\altaffilmark{1},
  C.-Y.~Ng\altaffilmark{2},
  B.~M.~Gaensler\altaffilmark{2},
  S.~M.~Ransom\altaffilmark{3},
  S.~Chatterjee\altaffilmark{4},
  J.~Reynolds\altaffilmark{5},
  and J.~Sarkissian\altaffilmark{5}
}

\altaffiltext{1}{Columbia Astrophysics Laboratory, Columbia University,
  New York, NY~10027, USA}
\altaffiltext{2}{Sydney Institute for Astronomy, School of Physics,
  The University of Sydney, NSW~2006, Australia}
\altaffiltext{3}{National Radio Astronomy Observatory, Charlottesville, 
  VA~22903, USA}
\altaffiltext{4}{Department of Astronomy, Cornell University, Ithaca,
  NY~14853, USA}
\altaffiltext{5}{CSIRO Parkes Observatory, Parkes, NSW~2870, Australia}

\begin{abstract}
The faint radio supernova remnant SNR~\snr\ is notable for a long and
thin trail that extends outward perpendicular from the edge of its
approximately circular shell.  In a search with the Parkes telescope
we have found a young and energetic pulsar that is located at the tip
of this collimated linear structure.  \psr\ has period $P=61$\,ms,
characteristic age $\tau_c = P/(2 \dot P) = 114$\,kyr, and spin-down
luminosity $\dot E = 1.4\times10^{36}$\,\ergs.  It is very faint, with
a flux density at 1.4\,GHz of about $75\,\mu$Jy.  From its dispersion
measure of 549\,\pccc, we infer $d \approx 8$\,kpc.  At this distance
and for an age comparable to $\tau_c$, the implied pulsar velocity in
the plane of the sky is $V_t = 300$\,\kms\ for a birth at the center of
the SNR, although it is possible that the SNR/pulsar system is younger
than $\tau_c$ and that $V_t > 300$\,\kms.  The highly collimated linear
feature is evidently the pulsar wind trail left from the supersonic
passage of \psr\ through the interstellar medium surrounding SNR~\snr.

\end{abstract}


\section{Introduction} \label{sec:intro} 

Pulsars convert their prodigious rotational energy into relativistic
winds of particles and fields.  When significantly confined by ambient
pressure, a portion of this outflow, or of shocked surrounding medium, may
be detected as a pulsar wind nebula (PWN).  PWNe are observed across the
electromagnetic spectrum in enormous diversity, reflecting differences in,
and serving as probes of, the properties of the pulsars and their winds,
and of their environments \cite[for a review, see][]{gs06}.

After pulsars escape the confines of their expanding supernova remnant
(SNR) shells, into media with relatively small sound speed, they typically
move supersonically.  Confinement by ram pressure of the ISM leads to
characteristically elongated bow shock PWNe.  These are observed at radio
or X-ray wavelengths, when the shocked pulsar wind emits synchrotron
radiation, and/or in the H$\alpha$ emission of collisionally excited ISM.
Among some 15 known pulsar bow shock nebulae, two-thirds are in the former
category \cite[see][]{cc02}, of which the best-studied example is the
``Mouse'' \citep{gvc+04}.  Given the variety of pulsar properties and
environmental conditions, it is desirable to increase the small sample
of pulsar bow shock nebulae available for detailed study.

SNR~\snr\ is a little-studied faint radio shell, approximately $15'$
in diameter and incomplete in the north-east, with a peculiar
collimated jet-like protrusion extending radially outward from
the north-west quadrant of the shell for approximately $8'$
\citep{kcr+87}.  Figure~\ref{fig:snr} (left) shows an 843\,MHz
radio image of SNR~\snr, from the Molonglo Observatory Synthesis
Telescope (MOST) survey of \citet{wg96}\footnote{Obtained from
http://www.physics.usyd.edu.au/astrop/wg96cat/msc.a.html.}.  Whiteoak \&
Green \nocite{wg96} noted that the protruding linear feature does not
have an infrared counterpart in $60\,\mu$m {\em IRAS}\ data, suggesting
that the radio emission from this source is non-thermal.  The overall
SNR morphology is reminiscent of a frying pan.

Such long collimated structures are rare \citep[see, e.g.,][]{ggm98,kp08},
and we considered it likely that this one could be generated by the
trailing wind of a fast-moving pulsar.  In this Letter we report the
discovery of \psr, which is highly likely to be responsible for this
remarkable feature and to be associated with SNR~\snr.

\section{Observations and Results} \label{sec:obs}

\begin{figure*}[t]
\centerline{
\hfill
\psfig{figure=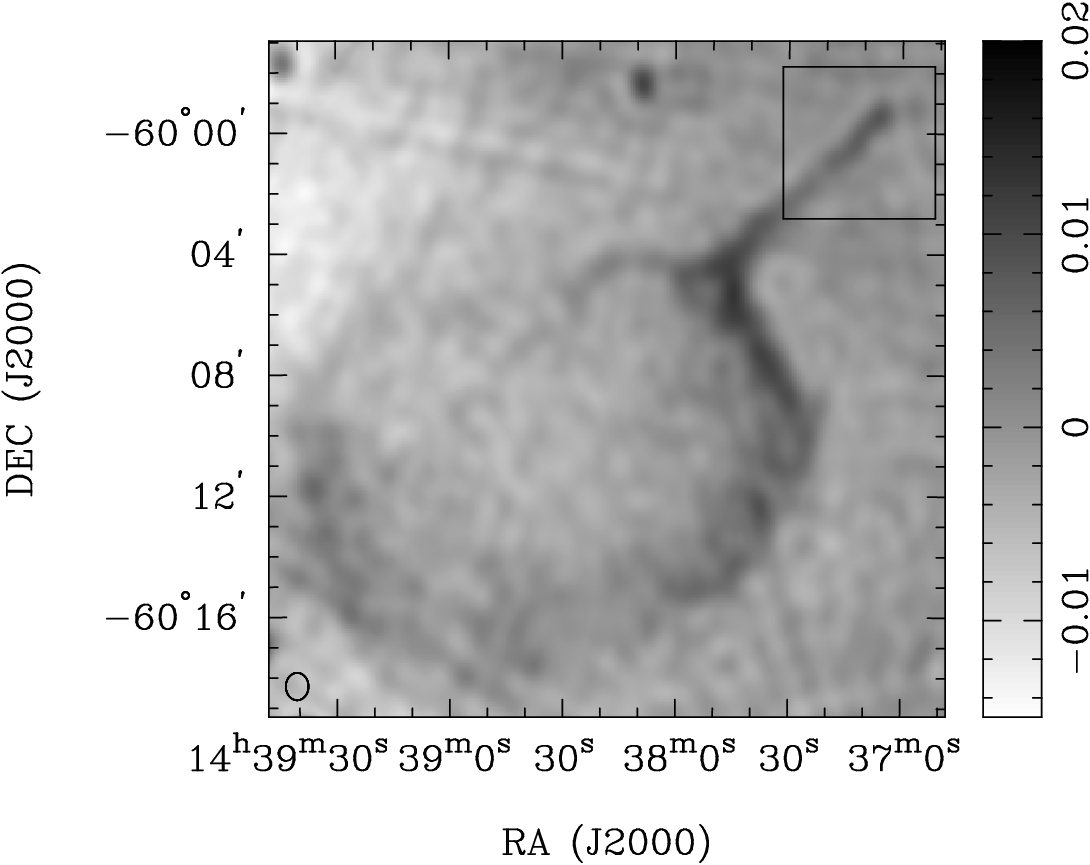,width=0.49\linewidth,angle=0}
\hfill
\psfig{figure=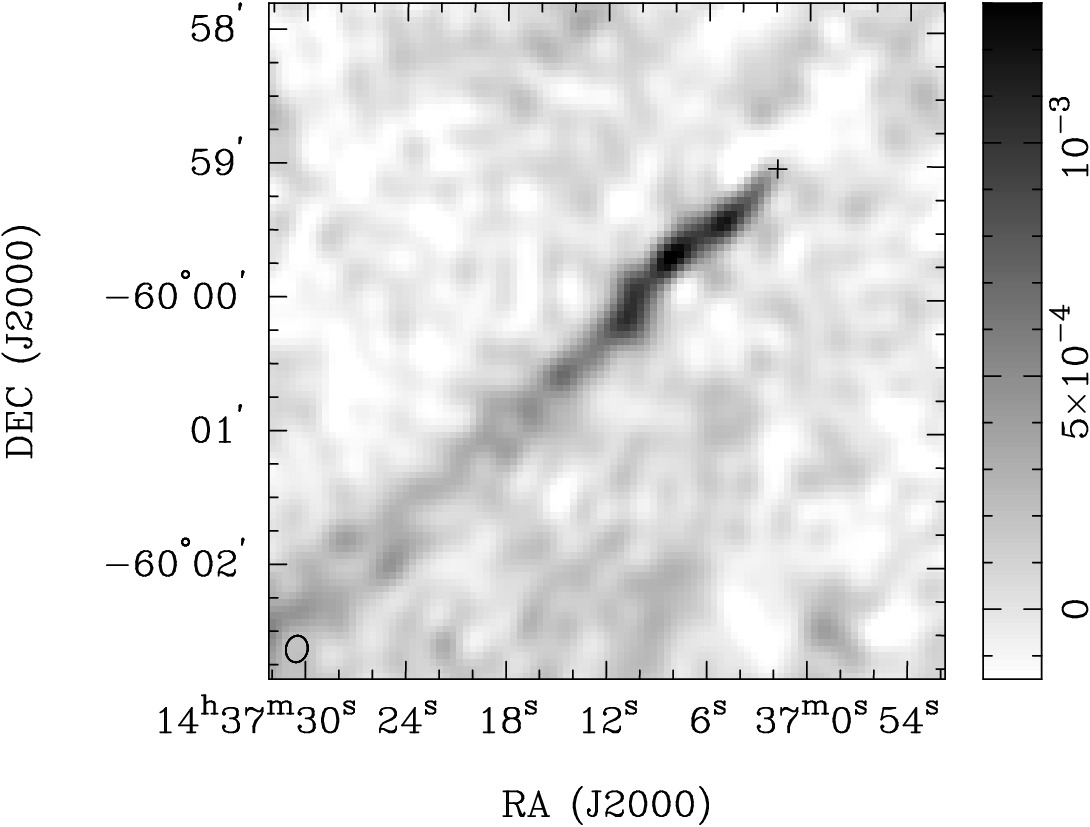,width=0.49\linewidth,angle=0}
\hfill
}
\caption{\label{fig:snr}
Left: MOST image of SNR~\snr\ at 843\,MHz \citep[data from][]{wg96}.
The beam size (shown at lower left) is $54''\times45''$, the rms noise
is 1.1\,mJy\,beam$^{-1}$, and the grey scale is linear, ranging from
--15\,mJy\,beam$^{-1}$ to 20\,mJy\,beam$^{-1}$.  The area bounded by
a square is shown on the right panel at higher resolution.  Right:
ATCA 2.4\,GHz image of \snr, zoomed in near the tip of the linear
trail, with the position of \psr\ marked by the cross.  The pulsar
positional uncertainty is $\sim 10$ times smaller than the cross (see
Table~\ref{tab:parms}).  The image has beam size $12''\times10''$,
rms of 0.13\,mJy\,beam$^{-1}$, and a linear scale ranging between
--0.15\,mJy\,beam$^{-1}$ and 1.3\,mJy\,beam$^{-1}$.
}
\end{figure*}

\begin{figure}[t]
\begin{center}
\includegraphics[angle=270,scale=0.45]{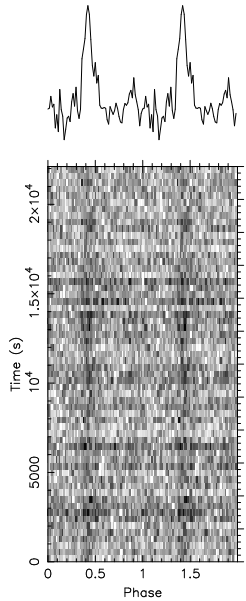}
\caption{\label{fig:psr}
Discovery observation of \psr\ at the Parkes telescope at 1.4\,GHz.
The pulse profile (shown twice; $P=61$\,ms) is displayed as a function
of time in the grey scale and summed at the top.
}
\end{center}
\end{figure}

We observed the tip of the linear feature in SNR~\snr\ (see left
panel of Figure~\ref{fig:snr}) with the ATNF Parkes telescope on 2008
June 17.  The 6.1\,hr observation used the central beam of a multibeam
receiver at a frequency of 1374\,MHz, spanning a bandwidth of 288\,MHz,
with the total radio power sampled in each of 96 polarization-summed
channels every 1\,ms and recorded to disk.  The data were searched for
periodic dispersed signals using standard techniques implemented in
PRESTO \citep{ran01,rem02}, similarly to other deep searches for young
pulsars \citep[e.g.,][]{crg+06}.  We identified in these data a pulsar
with period $P=61$\,ms and dispersion measure $\mbox{DM} = 549$\,\pccc\
(Figure~\ref{fig:psr}).  The distance is estimated from the DM and
the \citet{cl02} NE2001 electron density model to be 8\,kpc, which we
parameterize by $d_8 = d/(8\,\mbox{kpc})$.  We attempted to obtain a
precise position using a pulsar-gated observation with the Australia
Telescope Compact Array (ATCA) on 2008 July 23 (project CX156), but the
pulsar was too faint to be detected.

We have made regular timing observations of \psr\ at
Parkes since discovery.  Typically we observe the pulsar for
approximately 1\,hr at a frequency of 1.4\,GHz.  From each such
observation we obtain one time-of-arrival (TOA).  We have used
TEMPO\footnote{http://www.atnf.csiro.au/research/pulsar/tempo} and 23
TOAs spanning 1\,yr to obtain a phase-connected timing solution that
yields high-precision measurements of $P$, $\dot P$, R.A., and decl.\
(see Table~\ref{tab:parms}, where we also gather some important derived
parameters).  The residuals from this fit are featureless, and any
unmodeled rotational instability contributes a systematic uncertainty
to celestial coordinates of no more than $\sim 1''$.

\begin{deluxetable}{ll}
\tablewidth{0.95\linewidth}
\tablecaption{\label{tab:parms}Measured and Derived Parameters for \psr\ }
\tablecolumns{2}
\tablehead{
\colhead{Parameter} &
\colhead{Value}
}
\startdata
Right ascension, R.A. (J2000)\dotfill      & $14^{\rm h}37^{\rm m}01\fs91(3)$ \\
Declination, decl.\ (J2000)\dotfill        & $-59\arcdeg59'01\farcs4(3)$      \\
Galactic longitude, $l$ (deg)\dotfill      & 315.783                          \\
Galactic latitude, $b$ (deg)\dotfill       & 0.227                            \\
Spin period, $P$ (s)\dotfill               & 0.0616961234035(9)               \\
Period derivative, $\dot P$\dotfill        & $8.5870(5)\times 10^{-15}$       \\
Epoch (MJD)\dotfill                        & 54827.0                          \\
Post-fit rms timing residual (ms)\dotfill  & 0.7                              \\
Data span (MJD)\dotfill                    & 54634--55018                     \\
Dispersion measure, DM (\pccc)\dotfill                 & $549.6(11)$          \\
Rotation measure, RM (rad\,m$^{-2}$)\dotfill           & $-700\pm25$          \\
Flux density at 1.4\,GHz, $S_{1.4}$ ($\mu$Jy)\dotfill  & $\approx 75$         \\
Pulse FWHM ($P$)\dotfill                               & 0.11                 \\
Characteristic age, $\tau_c$ (kyr)\dotfill             & 114                  \\
Spin-down luminosity, $\dot E$ (\ergs)\dotfill         & $1.4\times10^{36}$   \\
Surface dipole magnetic field strength (Gauss)\dotfill & $7.4\times 10^{11}$  \\
Distance, $d$ (kpc)\tablenotemark{a}\dotfill           & $\approx 8$          \\
Radio luminosity, $S_{1.4} d^2$ (mJy\,kpc$^2$)\dotfill & $\sim 5$
\enddata
\tablecomments{Values in parentheses represent $1\,\sigma$ TEMPO uncertainties.}
\tablenotetext{a}{Distance is inferred from the DM and the \citet{cl02} model.}
\end{deluxetable}

We also made one polarimetric observation, using a digital filterbank
to record all four Stokes parameters for 5.8\,hr on 2008 September 27.
These data were analyzed with PSRCHIVE \citep{hvm04}.  The emission
is approximately 50\% linearly polarized throughout the pulse profile,
with a fairly flat position angle of linear polarization that does not
constrain the geometry.  Faraday rotation is measured to be $\mbox{RM} =
(-700\pm25)$\,rad\,m$^{-2}$; along with the DM, this implies an average
Galactic magnetic field along the line of sight, weighted by the local
free electron density, of $1.6\,\mu$G, directed away from us.  The pulse
profile shows a hint of an interpulse (see Figure~\ref{fig:psr}), while
the main pulse appears slightly asymmetric.  The possible one-sided
exponential tail of width $\approx 5$\,ms could be caused by multipath
propagation through the ISM (this is half the value expected for this
location and DM from the NE2001 model).  From this flux-calibrated
observation we measure a 1.4\,GHz period-averaged flux density of $S_{1.4}
\approx 75\,\mu$Jy.

SNR~\snr\ has been observed with the ATCA at frequencies of 1.4\,GHz
and 2.4\,GHz.  We have reprocessed the archival 2.4\,GHz data taken
on 1998 July 5, 1999 March 25 and April 20 with the 750E, 1.5B and
1.5C array configurations, respectively. Each observation had 12\,hr
integration time, and PKS~1934--63 and PKS~1251--71 were observed
as the primary and secondary calibrators to determine the flux
density and antenna gains.  Our data reduction was carried out in
MIRIAD\footnote{http://www.atnf.csiro.au/computing/software/miriad}.
We excluded any baselines longer than 2\,km to provide a uniform $u$-$v$
sampling and to boost the signal-to-noise ratio. After flagging the
bad data points and scans during poor atmospheric phase stability, we
applied the multifrequency synthesis technique with uniform weighting
to form the images. This can improve the $u$-$v$ coverage and minimize
the sidelobes.  The resulting image was then deconvolved using a maximum
entropy algorithm \citep{ssb96} and restored with a Gaussian beam of
FWHM $12\arcsec \times 10\arcsec$. Figure~\ref{fig:snr} (right) shows
the image zoomed in near the pulsar position.

The position of \psr\ was observed with the \swift\ X-ray
telescope (in PC mode) on 2008 July 5 for a total of 9.6\,ks.
We detect only one photon within $5''$ of the pulsar position,
and consider five photons to represent a flux upper limit.  Using
PIMMS\footnote{http://heasarc.gsfc.nasa.gov/Tools/w3pimms.html} with
the total Galactic neutral hydrogen column density in this direction,
$N_H = 1.6\times10^{22}$\,cm$^{-2}$, and a typical power-law model with
photon index $\Gamma = 1.5$, the unabsorbed flux in the 2--10\,keV range
is $f_X < 4.5\times10^{-14}$\,erg\,cm$^{-2}$\,s$^{-1}$.  The resulting
X-ray luminosity is $L_X < 3.5\times10^{32}\,d_8^2$\,\ergs, or $L_X/\dot
E < 2.5\times10^{-4}\,d_8^2$.  This limit is somewhat below the ``usual''
X-ray luminosities from pulsars (or their PWNe) with comparable $\dot E$,
which however have a large spread \citep[see, e.g.,][]{pccm02}.

\section{Discussion} \label{sec:disc} 

We have discovered \psr, a young pulsar with $\tau_c = 114$\,kyr that
is located precisely at the tip of an unusual collimated linear feature
(right panel of Figure~\ref{fig:snr}), suggesting a physical association.
In turn the linear feature seamlessly connects with the shell SNR~\snr,
which is brightest in the area near the point of apparent contact
(left panel of Figure~\ref{fig:snr}), suggesting a physical connection.
By analogy with well-studied examples of pulsars and their bow shock
PWNe \citep[e.g.,][]{gvc+04}, we infer that the long non-thermal nebula
trailing \psr\ is caused by its shocked relativistic wind as the pulsar
moves supersonically through the local ISM and away from SNR~\snr.

The pulsar/SNR association is located in the direction of the Crux
spiral arm.  At a distance of 8\,kpc \citep[according to the NE2001 model
of][]{cl02}, this system is on the inside edge of the arm.  The $\mbox{RM}
= -700$\,rad\,m$^{-2}$ measured for \psr\ is typical for this part of the
Galaxy \citep{bhg+07}.  Many high-DM pulsars known within the Crux arm
have $-900 < \mbox{RM}/(\mbox{rad\,m$^{-2}$}) < -500$, and are located
at $d \sim 6$\,kpc \citep{hml+06}.  If the NE2001 model Crux arm has a
smaller electron density than the real arm or does not extend as far in
width, it is entirely possible that the true pulsar distance is as small
as 6\,kpc.  We consider it reasonable that the NE2001 fractional distance
uncertainty for this system is about 25\%, i.e., $d_8 = 1\pm0.25$.

\psr\ is located $15'$ ($=35\,d_8$\,pc) from the geometric center of
the shell \snr, its putative place of birth.  For a true age $\tau$,
parameterized as $\tau_{114} = \tau/\tau_c$, the pulsar velocity
in the plane of the sky is $V_t = 300\,d_8/\tau_{114}$\,\kms,
which is a perfectly reasonable velocity for a young neutron star
\citep[e.g.,][]{acc02}.  For almost any ISM phase, this is also a highly
supersonic velocity.

While $\approx 300$\,\kms\ is a plausible velocity for a young pulsar,
we now consider the possibility that for \psr\ the true velocity may be
substantially larger.  We know little about the ISM density near SNR~\snr,
but it seems reasonable to question whether the age of the SNR really is
as large as 114\,kyr, especially in view of its relatively undisturbed,
if incomplete, circular symmetry.  Also, the SNR shell radius, $R_{\rm
SNR} = 17\,d_8$\,pc, may be small for such a large age: for adiabatic
expansion, the Sedov solution implies $\tau \approx 22\,d_8^{5/2}
(n_0/E_{51})^{1/2}$\,kyr, for a hydrogen ambient medium of density
$n_0$\,cm$^{-3}$ and SN explosion kinetic energy of $10^{51}\,E_{51}$\,erg
(a search for thermal X-rays from the SNR could provide useful constraints
on its age).  In addition, for the nominal age, the long pulsar wind trail
(of length $17\,d_8$\,pc) must have remained remarkably collimated for
about 50\,kyr.  A detailed study of the PWN (now underway using the ATCA)
may find whether this is reasonable.  If on the other hand $\tau_{114}
\ll 1$, these possible difficulties disappear.  In that case, the
pulsar age is $\ll \tau_c$.  For instance, if $\tau = 0.2\,\tau_c$, the
implied initial spin period of the neutron star is $P_0 \approx 55$\,ms.
Such relatively large initial periods, close to their current periods,
are inferred for other pulsars \citep[e.g.,][]{crg+06,krv+01}, and this
could be the case as well for \psr.  And if $\tau_{114} \approx 0.2$,
then $V_t \approx 1500$\,\kms.  The expected proper motion for the
pulsar (or, as its proxy, for the head of the bow shock nebula) is $\mu
= 8/\tau_{114}$\,mas\,yr$^{-1}$.  If indeed $\tau_{114} \ll 1$, this
could be measurable with the ATCA.  Thus, multiwavelength observations
of SNR~\snr, its associated \psr, and its bow shock PWN, may usefully
constrain the properties of this newly identified system, and put it
into context in the zoo of young neutron stars and their environments.

\begin{figure}[t]
\begin{center}
\includegraphics[angle=0,scale=0.40]{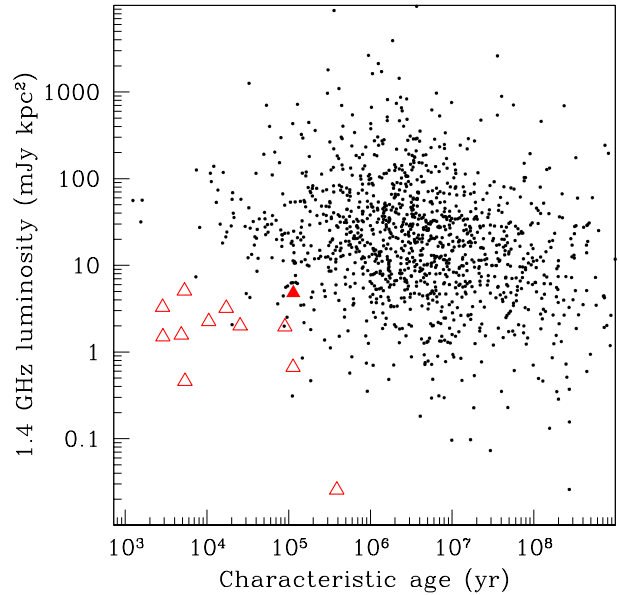}
\caption{\label{fig:lrtau}
Radio luminosity at 1.4\,GHz ($L_{1.4}$) vs.\ characteristic age
($\tau_c$) for 1259 pulsars in the Galactic disk with $\tau_c <
10^9$\,yr.  The pulsars denoted by red triangles (\psr\ is the filled
one) were discovered in directed searches of PWNe/energetic targets
(see \S~\ref{sec:disc}).  All other information is from the ATNF pulsar
catalog \citep{mhth05}.  }
\end{center}
\end{figure}

\psr\ has a very small flux density, and it is not surprising that
it was not identified previously in relatively shallow undirected
surveys.  Its location was observed in the Parkes multibeam survey
of the Galactic plane \citep[e.g.,][]{mlc+01}, and although we do
detect the pulsar faintly in archival data from 1997 October 26, it
is not significant in an unbiased search.  Its luminosity, $L_{1.4}
\equiv S_{1.4} d^2 \approx 5\,d_8^2$\,\rlum, is low by historical
standards, but following numerous young pulsar discoveries in
recent years is now unremarkable.  We update the current state of
play in Figure~\ref{fig:lrtau} (using data from the ATNF pulsar
catalog\footnote{http://www.atnf.csiro.au/research/pulsar/psrcat}
along with revised and unpublished fluxes in some cases),
where we highlight the radio pulsars discovered in deep
directed searches of PWNe/high-energy targets \citep[][and this
work]{clb+02,cmg+02,cmgl02,csl+02,crg+06,crgl09,crr+09,gmga05,hcg+01,kjk+08,rhr+02}.
The most sensitive searches possible with Parkes, in the absence of
significant scattering \citep[see][]{crgl09,ojk+08}, reach threshold
$L_{1.4} \sim 1$\,\rlum\ at $d \sim 5$\,kpc.  Figure~\ref{fig:lrtau}
then suggests that while the faint tail of the population remains out of
reach in a significant Galactic volume (even at more sensitive telescopes
like the GBT), there is also a substantial population of young pulsars
with $L_{1.4} \sim 5$\,\rlum\ that is detectable to great distances,
given a maximum effort on carefully selected targets, exemplified here
by \psr\ and SNR~\snr.

\acknowledgements
We thank Matthew Young and Dick Manchester for doing the original ATCA
observations.  We thank Phil Edwards for scheduling the pulsar-gated
observation at the ATCA, and the \swift\ project scientist and staff
for the \swift\ observation.  We are grateful to Simon Johnston for
discussions concerning the polarimetric observation, and to Jules Halpern
for help with the \swift\ data.  The Parkes Observatory and the ATCA are
part of the Australia Telescope, which is funded by the Commonwealth
of Australia for operation as a National Facility managed by CSIRO.
The MOST is operated by The University of Sydney with support from the
Australian Research Council and the Science Foundation for Physics within
The University of Sydney.  This work was supported in part by the NSF
through grant AST-0908386 to F.C.  C.-Y.N.\ and B.M.G.\ acknowledge the
support of Australian Research Council grant FF0561298.

{\em Facilities:} \facility{ATCA}, \facility{Molonglo}, \facility{Parkes
(PDFB, PMDAQ)}, \facility{Swift (XRT)}

\end{document}